\documentclass[3p,times,procedia]{elsarticle}
\usepackage{nupha_ecrc}
\usepackage{xspace}

\volume{00}

\firstpage{1}

\journalname{Nuclear Physics A}

\runauth{J.D. Osborn for the PHENIX collaboration}

\jid{nupha}

\jnltitlelogo{Nuclear Physics A}

\usepackage{amssymb}

 \usepackage{lineno}

\usepackage[figuresright]{rotating}

\newcommand{\pp}{$p$$+$$p$\xspace}
\newcommand{\pion}{$\pi^0$\xspace}
\newcommand{\pt}{$p_T$\xspace}
\newcommand{\zt}{$z_T$\xspace}
\newcommand{\sqsn}{$\sqrt{s_{NN}}$\xspace}
\newcommand{\sqs}{\mbox{$\sqrt{s}$}\xspace}
\newcommand{\pout}{\mbox{$p_{out}$}\xspace}
\newcommand{\ptassoc}{\mbox{$p_T^{\rm assoc}$}\xspace}
\newcommand{\pttrig}{\mbox{$p_T^{\rm trig}$}\xspace}
\newcommand{\dphi}{\mbox{$\Delta\phi$}\xspace}
\newcommand{\xe}{\mbox{$x_E$}\xspace}

\newcommand{\gevc}{GeV/$c$\xspace}

\begin{document}

\begin{frontmatter}



\dochead{XXVIIth International Conference on Ultrarelativistic Nucleus-Nucleus Collisions\\ (Quark Matter 2018)}

\title{PHENIX results on jet modification with $\pi^0$- and photon-triggered two particle correlations in \pp, $p(d)$+Au, and Au+Au collisions}


\author{J.D. Osborn for the PHENIX Collaboration}

\address{}

\begin{abstract}
As a colorless probe, direct photons balance the \pt of the away-side jet at 
leading order. Direct photon-hadron correlations are thus an excellent probe for
nuclear structure and QCD effects, including parton energy loss in the 
Quark-Gluon Plasma. PHENIX has measured \pion and direct photon-triggered two-particle azimuthal 
correlations in a variety of collision systems ranging from \pp to Au+Au at 200 
GeV. In $p$$+$Au and $d$$+$Au collisions, no modification of the per-trigger jet yield or
away-side correlation width compared to \pp collisions is observed for direct 
photon triggered correlations while an increase in the away-side width for 
\pion triggered correlations in $p$$+$Au has been measured. In Au+Au collisions, direct photons have been identified statistically as well as using an isolation cut. Combining data sets from different collision systems allows us to quantify the transition from suppression at high 
\zt $(p_{T,h}/p_{T,\gamma})$ to the enhancement of low \zt particles relative to \pp, and to study this transition as a function of trigger \pt. 
\end{abstract}


\end{frontmatter}


\section{Introduction}\label{Introduction}

Two particle correlations are a powerful tool to probe a variety of quantum chromodynamics (QCD) interactions. In particular, correlations can provide information about the $2\rightarrow 2$ hard partonic process in a given event. At the PHENIX experiment, high \pt correlations are generally measured with a leading \pion or direct photon. Dihadron correlations have the benefit of being a high statistics measurement; additionally, the leading \pion and away-side hadrons may both encounter strong force interactions leading to larger modifications. Nonetheless, direct photon correlations are particularly sought after due to the photon emerging directly from the hard scattering; therefore, they provide the most precise measure of the initial partonic hard scattering kinematics in hadronic collisions where final-state hadrons are measured. Photons additionally do not interact strongly, and thus are insensitive to QCD effects such as color exchanges or medium interactions. Therefore, the direct photon provides an excellent benchmark to understand these interactions. 

The PHENIX collaboration has recently measured high \pt two-particle angular correlations in a variety of collision systems. Direct photon-hadron correlations have been measured in $d$+Au and Au+Au collisions, while \pion-hadron correlations have been measured in $p$+Al and $p$+Au collisions at \sqsn=~200 GeV. Both \pion-hadron and direct photon-hadron correlations have also been measured in \pp collisions at \sqs=~200 and 510 GeV. The PHENIX detector is capable of measuring these correlations in azimuth, \dphi, utilizing its central arm electromagnetic calorimeters to measure trigger photons or neutral pions and tracking system, composed of drift and pad chambers, to measure associated hadrons. The central arms cover an azimuthal acceptance of approximately $\pi$ radians and a pseudorapidity acceptance of $|\eta|<0.35$.

\section{Studying effects from color flow in \pp collisions}\label{pp_results}

In the last several decades, there has been strong interest in the multi-dimensional structure of the proton~\cite{Aidala:2012mv}. In particular, this interest has led to new predictions that are specific to non-Abelian gauge invariant quantum field theories. For example, in a transverse-momentum-dependent (TMD) framework, where the nonperturbative transverse momentum of partons within a hadron are explicitly considered, QCD factorization breaking has been predicted in the process \pp$\rightarrow h_1$$+$$h_2$$+$$X$~\cite{Rogers:2010dm}. In hadronic collisions, where at least one final-state hadron is measured and a TMD framework is applicable, soft gluons may be exchanged in both the initial and final states. This leads to complex color flows that correlate the partons and result in the breakdown of TMD factorization.

\begin{figure}[tbh]
	\includegraphics[width=0.5\textwidth]{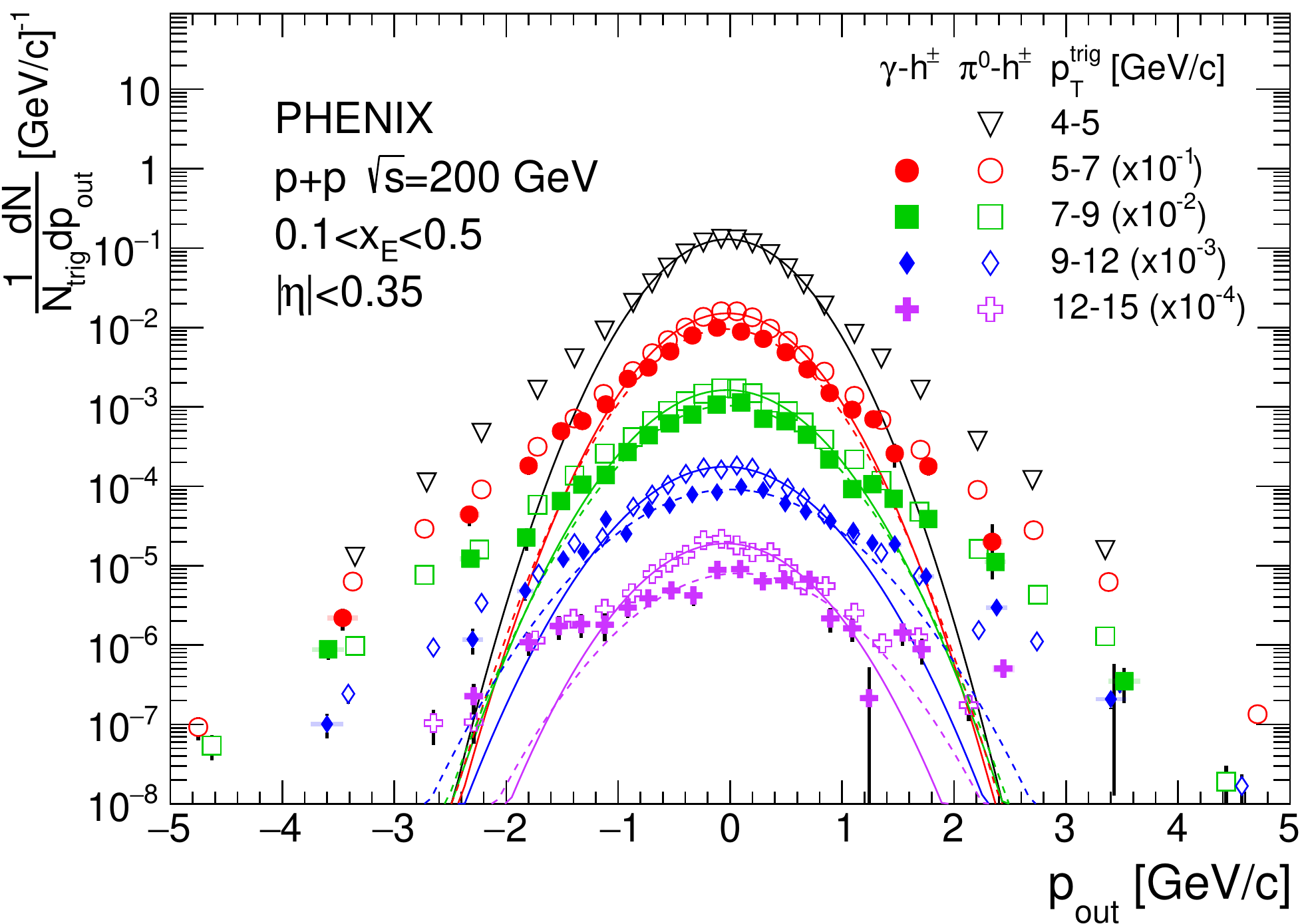}
	\includegraphics[width=0.5\textwidth]{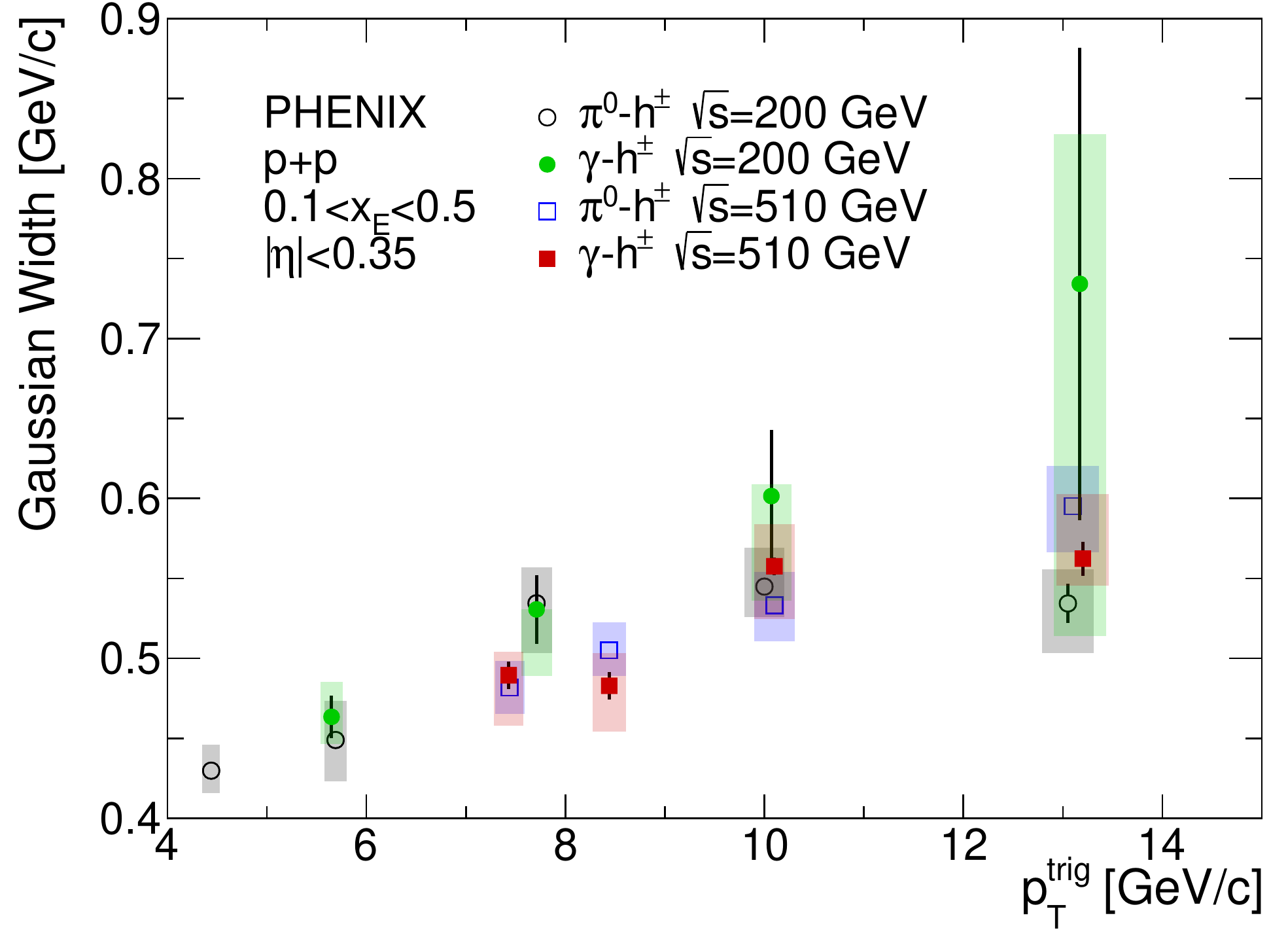}
	\caption{The measured \pout distributions in \pp collisions are shown for dihadron and direct photon-hadron correlations (left). The Gaussian widths of these distributions are shown as a function of \pttrig compared to the previous \sqs=~510 GeV results~\cite{Aidala:2018bjf} (right).}
	\label{fig:pouts}
\end{figure}

To study these predicted effects, PHENIX has measured the $\pout=\ptassoc\sin\dphi$ distributions in a fixed bin of $\xe=-(\ptassoc\cdot\pttrig)/(|\pttrig|^2)$~\cite{Aidala:2018bjf}. Constructing the correlations in a fixed bin of \xe allows for a better comparison between bins of \pttrig since \xe is a proxy for the away-side longitudinal momentum fraction $z$. The measured \pout distributions are shown in the left panel of Fig.~\ref{fig:pouts}, where the open (filled) markers are the \pion-hadron (direct photon-hadron) correlations and each marker color shows a different \pttrig bin. The distributions show a two component shape, which is characterized by a Gaussian fit at small \pout and a power law-like tail from the Gaussian at large \pout. The two shapes arise due to sensitivity to soft gluon radiation in the $|\pout|\sim0$ region and hard gluon radiation in the $|\pout|\gg0$ region, indicating the applicability of a TMD framework since there is sensitivity to both a hard and soft transverse momentum scale. The Gaussian widths, shown in the right panel of Fig.~\ref{fig:pouts}, increase as a function of \pttrig, which is qualitatively consistent with nonperturbative momentum width measurements in the Drell-Yan process where factorization is predicted to hold (see e.g.~\cite{Landry:2002ix}). Future quantitative phenomenological comparisons will be necessary which compare these different processes to understand the quantitative magnitude of TMD factorization breaking effects.

\section{Nonperturbative transverse momentum broadening in $p$+A}

In proton-nucleus collisions, dihadron correlations can be used to search for QCD interactions with the nucleus. In this case, the near-side triggered \pion is more likely to emerge near the surface of the nucleus and thus the away-side hadron may on average experience a larger path length than in direct photon-hadron correlations; therefore, when searching for nuclear effects, dihadron correlations may be more sensitive to modification than direct photon-hadron correlations. PHENIX has measured the \pout distributions on the away-side and near-side in \pp, $p$+Al, and $p$+Au collisions. Similarly to the previous \pp results, the Gaussian widths are extracted from each collision species, and the squared width differences are determined to search for transverse momentum broadening in $p$+A collisions.

\begin{figure}[tbh]
	\includegraphics[width=0.5\textwidth]{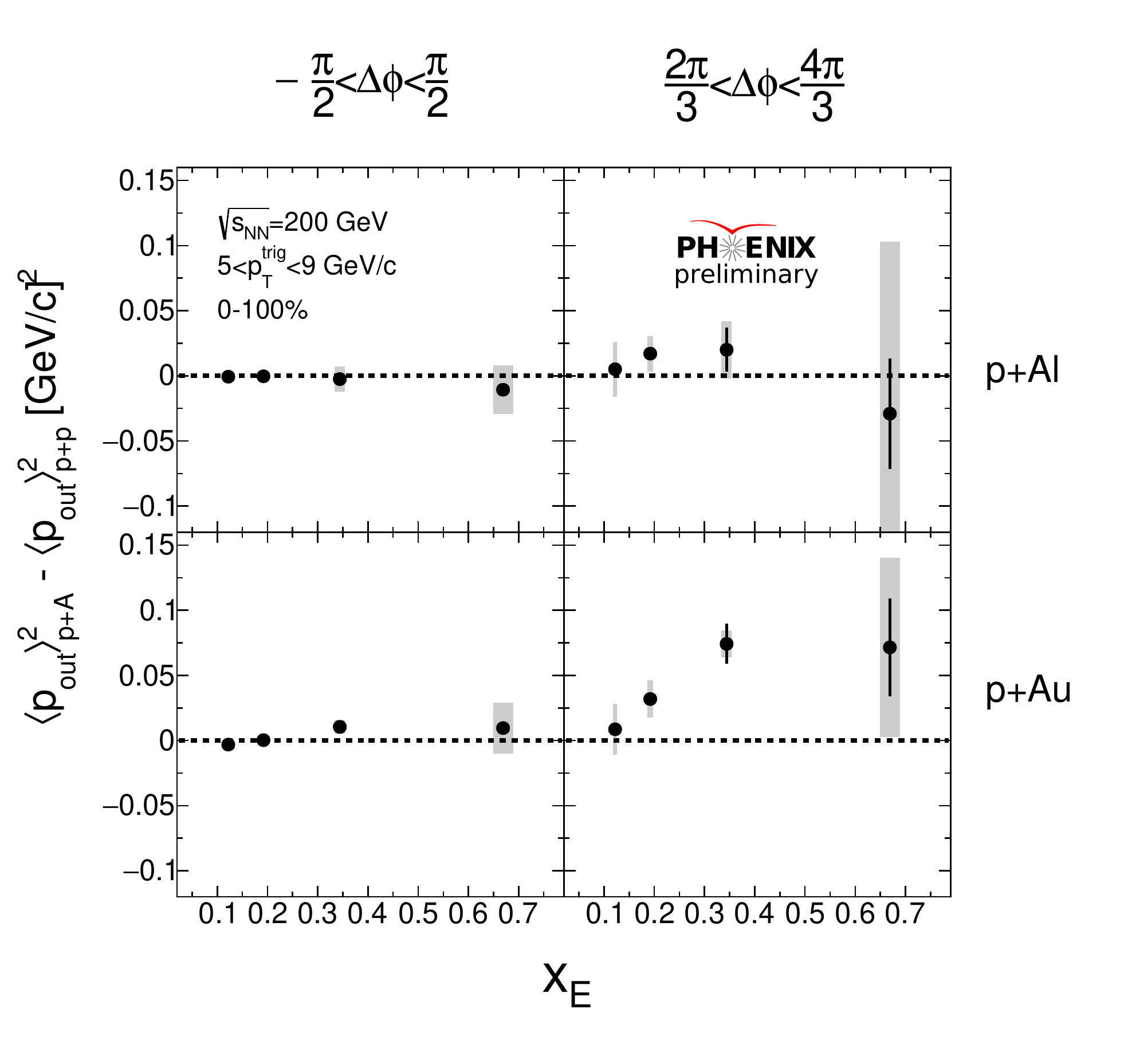}
	\includegraphics[width=0.5\textwidth]{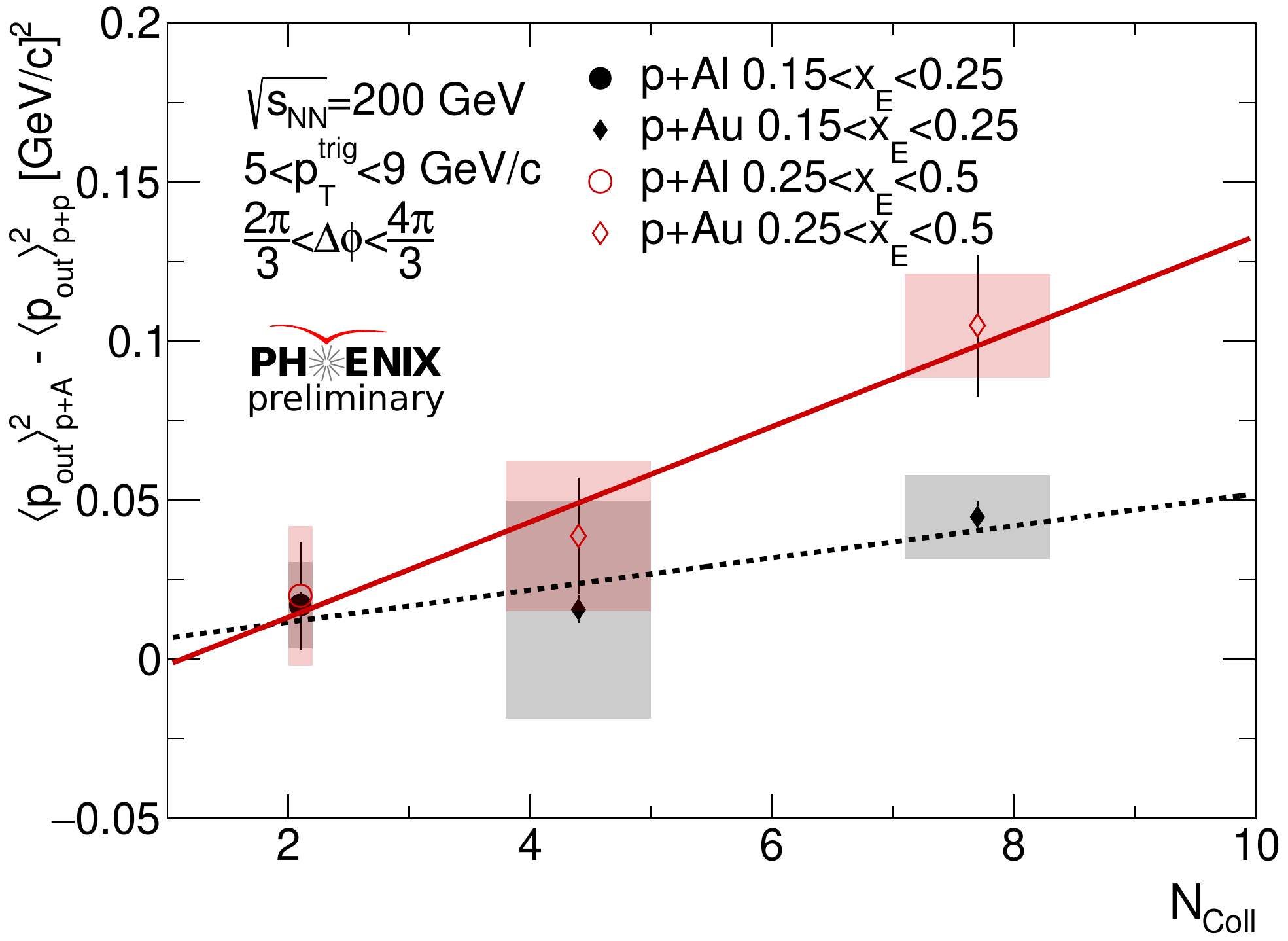}
	\caption{The squared Gaussian width differences between $p$+Al-\pp and $p$+Au-\pp are shown on the near and away sides (left). The away-side transverse momentum broadening is shown as a function of $N_{\rm coll}$ (right).}
	\label{fig:width_diff}
\end{figure}

The squared Gaussian width differences between $p$+A and \pp are shown in the left panel of Fig.~\ref{fig:width_diff} as a function of \xe. The top (bottom) row shows the width differences between $p$+Al and \pp ($p$+Au and \pp), while the left (right) column shows the near-side (away-side) width differences. The left column shows that, within the current uncertainties, there is no difference between the near-side nonperturbative widths in $p$+Al(Au) and \pp for the \xe range measured. The lack of near-side transverse momentum broadening suggests that final-state fragmentation differences in \pp and $p$+A collisions are small. In the right column, there is a nonzero and statistically significant nonperturbative momentum width difference between $p$+Au and \pp; a small difference can be seen in the away-side width differences between $p$+Al and \pp, however the uncertainties are consistent with no difference. To study this nonperturbative transverse momentum broadening as a function of collision system size, the two \xe bins where a nonzero difference can be seen are shown as a function of $N_{\rm coll}$ in the right panel of Fig.~\ref{fig:width_diff}. The broadening is a function of $N_{\rm coll}$, indicated by the linear fits to each \xe bin. This dependence may set tight constraints on energy loss within a nucleus; however, the correlations may also be sensitive to additional initial-state partonic transverse momentum in a nucleus in addition to ``Cronin'' mechanisms.

\section{Isolation cut method for direct photons in Au+Au}

Since direct photons are expected to be produced with little activity in the near vicinity, isolation cuts are often used to boost their signal-to-background from decay photon processes. The implementation of these cuts is common in \pp and $p(d)$+Au collisions; however, their use in Au+Au collisions can be challenging due to the larger underlying-event backgrounds. PHENIX has, for the first time at RHIC, measured isolated direct photon-hadron correlations in Au+Au collisions using a modified statistical subtraction method similarly to what is used in \pp collisions. The isolation cut increases the statistical significance of the direct photon yields in Au+Au collisions and and allows a more precise study of the \zt~=~\ptassoc/\pttrig dependence of away-side hadron suppression and enhancement. In these new results, direct photons and associated hadrons are collected in the range $5<\pttrig<15$ \gevc and $1<\ptassoc<10$ \gevc, respectively.

\begin{figure}[tbh]
	\centering
	\includegraphics[width=0.38\textwidth]{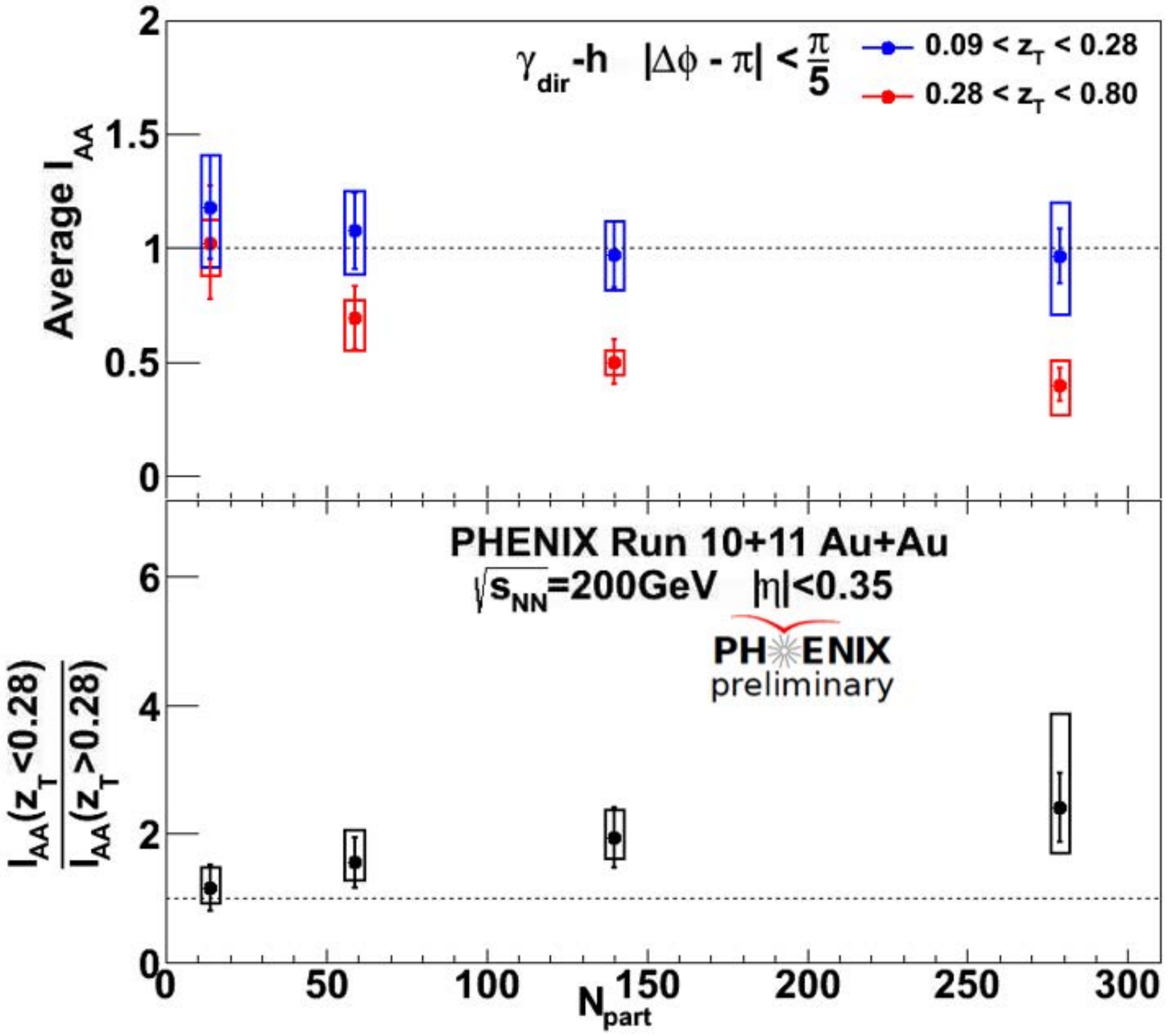}
	\includegraphics[width=0.5\textwidth]{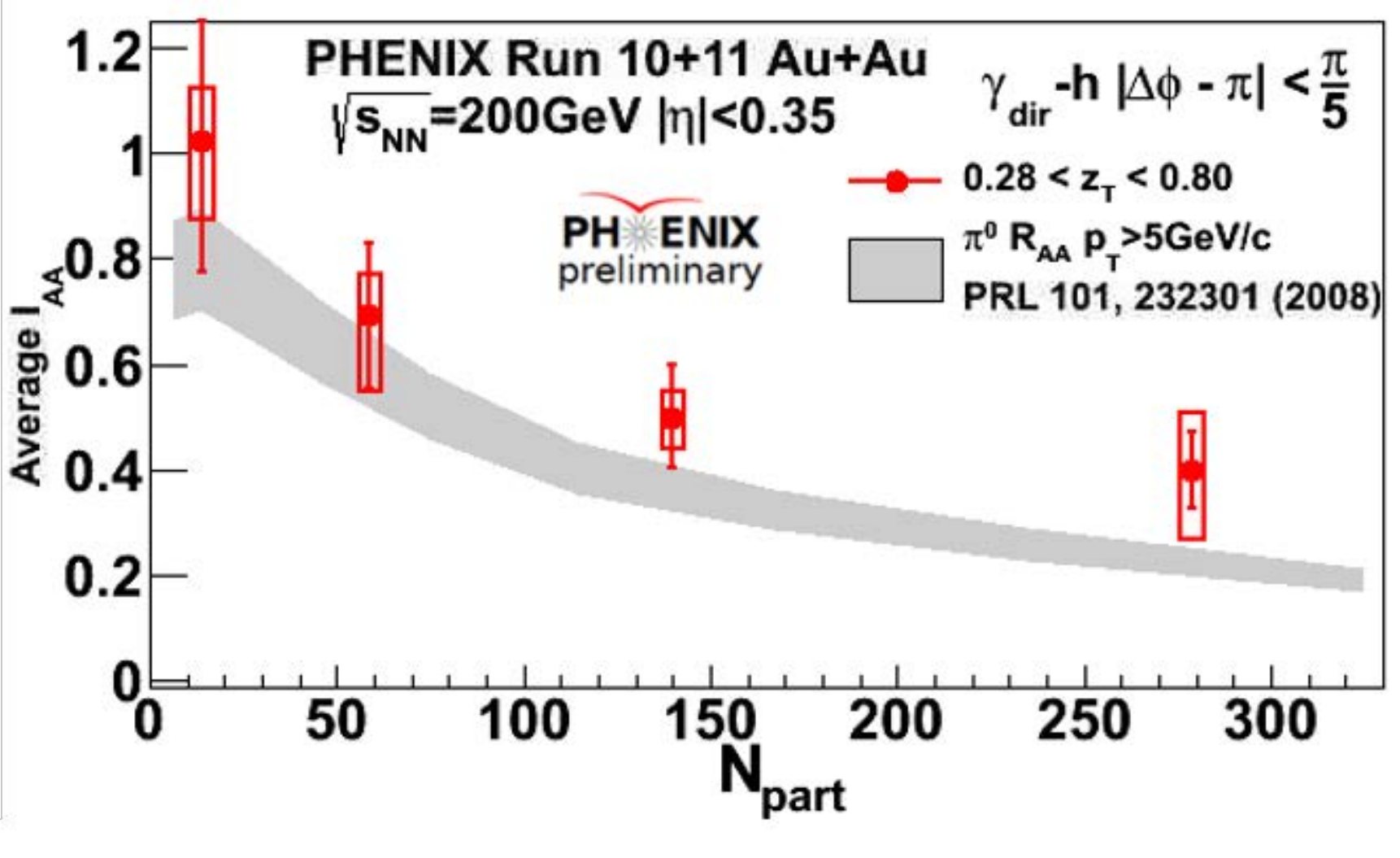}
	\caption{The measured $I_{AA}$ in isolated direct photon-hadron correlations in Au+Au and \pp collisions is shown as a function of centrality for high and low \zt particles (left). The high \zt isolated direct photon-hadron $I_{AA}$ is shown compared to the inclusive \pion $R_{AA}$ (right).}
	\label{fig:iaa}

\end{figure}

The low and high \zt away-side hadron $I_{AA}=Y_{AA}/Y_{pp}$, where $Y_{AA}$ is the integrated away-side yield in Au+Au or \pp collisions as indicated, as a function of centrality is shown in the left panel of Fig.~\ref{fig:iaa} for isolated direct photon-hadron correlations. The $I_{AA}$ values show the characteristic suppression at high \zt, in red, and enhancement relative to suppression at low \zt, in blue, indicative of quark gluon plasma formation. The larger accumulated data sample and isolation cut implementation allow for a more robust characterization compared to the previous PHENIX publication~\cite{Adare:2012qi} of this enhancement to suppression ratio, shown in the bottom of the left panel. In particular, this ratio shows a clear monotonic increase as a function of centrality, which is now more precisely identified. The high \zt suppression is also shown compared to the inclusive \pion $R_{AA}$ in the right panel of Fig.~\ref{fig:iaa}. The isolation cone implementation significantly improves the uncertainties from the previous publication~\cite{Adare:2009vd}.

\section{Summary}
The PHENIX collaboration has measured direct photon-hadron and dihadron correlations in \pp, $p$+Al, $p$+Au, and Au+Au collisions at \sqsn=~200 GeV. In \pp collisions, processes predicted to be sensitive to color flow in a TMD framework are studied and found to be qualitatively similar to Drell-Yan measurements. In $p$+A collisions, away-side transverse momentum broadening has been measured and this broadening appears to be dependent on $N_{\rm coll}$. In Au+Au collisions, an isolation cut has significantly improved the systematic uncertainties on direct photon-hadron correlations and thus the study of high and low associated hadron \zt suppression and enhancement, respectively. The analysis of the additional recent high statistics Au+Au data sets from PHENIX will continue to provide important constraints to our understanding of QCD. 






\bibliographystyle{elsarticle-num}
\bibliography{qm18}







\end{document}